\documentclass[prl,twocolumn,showpacs,preprintnumbers,amsmath,amssymb,superscriptaddress]{revtex4-1}
\usepackage{graphicx}
\usepackage{dcolumn}
\usepackage{bm}
\usepackage{color}

\begin{document}

\title{Closed loop approach to thermodynamics}
\author{C. Goupil}\email{christophe.goupil@univ-paris-diderot.fr}
\affiliation{Laboratoire Interdisciplinaire des Energies de Demain, LIED/CNRS UMR 8236 Universit\'e Paris Diderot; B\^at. Lamarck B 35 rue H\'el\`ene Brion 75013 Paris}
\author{H. Ouerdane}
\affiliation{Russian Quantum Center, 100 Novaya Street, Skolkovo, Moscow Region 143025, Russia}
\affiliation{UFR Langues Vivantes Etrang\`eres, Universit\'e de Caen Normandie, Esplanade de la Paix 14032 Caen, France}
\author{E. Herbert}
\affiliation{Laboratoire Interdisciplinaire des Energies de Demain, LIED/CNRS UMR 8236 Universit\'e Paris Diderot; B\^at. Lamarck B 35 rue H\'el\`ene Brion 75013 Paris}
\author{G. Benenti}
\affiliation{Center for Nonlinear and Complex Systems, Dipartimento di Scienza e Alta Tecnologia, Universit\`a degli Studi dell'Insubria, via Valleggio 11, 22100 Como, Italy} 
\affiliation{Istituto Nazionale di Fisica Nucleare, Sezione di Milano, via Celoria 16, 20133 Milano, Italy}
\author{Y. D'Angelo}
\affiliation{Laboratoire Interdisciplinaire des Energies de Demain, LIED/CNRS UMR 8236 Universit\'e Paris Diderot; B\^at. Lamarck B 35 rue H\'el\`ene Brion 75013 Paris}
\affiliation{Laboratory of Mathematics  J.A. Dieudonn\'e, CNRS UMR 7351 University of Nice-Sophia Antipolis Parc Valrose, Nice, France}
\author{Ph. Lecoeur}
\affiliation{Institut d'Electronique Fondamentale, Universit\'e Paris Sud CNRS, 91405 Orsay, France, CNRS, UMR 8622, 91405 Orsay, France}
\date{\today}

\begin{abstract}
We present the closed loop approach to linear nonequilibrium thermodynamics considering a generic heat engine dissipatively connected to two temperature baths. The system is usually quite generally characterized by two parameters: the output power $P$ and the conversion efficiency $\eta$, to which we add a third one, the working frequency $\omega$. We establish that a detailed understanding of the effects of the dissipative coupling on the energy conversion process, necessitates the knowledge of only two quantities: the system's feedback factor $\beta$ and its open-loop gain $A_{0}$, the product of which, $A_{0}\beta$, characterizes the interplay between the efficiency, the output power and the operating rate of the system. By placing thermodynamics analysis on a higher level of abstraction, the feedback loop approach provides a versatile and economical, hence a very efficient, tool for the study of \emph{any} conversion engine operation for which a feedback factor may be defined. 
\end{abstract}
\maketitle

\paragraph*{Introduction}

In the search for increased heat-to-work conversion efficiency, conversion engines have become quite complex systems, the detailed description of which, including their interactions with their environment, is daunting. Thermodynamics occupies a particular position in research, as a combination of feedback and feed forward between engineering and science, and offers the ideal framework for such purpose. Though classical, near-equilibrium and finite-time thermodynamics \cite{Callen,Onsager,deGroot,CurzonAhlborn,vdBroeck,Schmiedl,Tu,Apertet2}
are based on firm grounds, original \emph{compact} approaches able to analyze the behavior of a conversion engine --- and particularly the problem of conversion efficiency at maximum power under realistic coupling conditions to heat sources and sinks \cite{ApertetEPL,ApertetEPLr} --- are highly desirable. This is reflected by the abundant literature devoted to this class of problems (see Refs.~\cite{Andresen,Seifert,Ouerdane2015a} for recent reviews). Further, as thermodynamic systems may be mostly described using model-specific derivations, a comparison between them may prove quite intricate, and a generic common study of these is even lacking. The purpose of this paper is to provide the general framework that makes such a common description possible. 

The closed loop approach we present in this article essentially lies on two key assumptions: i) the basic equations governing the input and output of the system can be studied with diagrammatic reasoning; and ii) a feedback term for the power conversion, that is a function of the system working conditions, can be evidenced. In this article, we show how to construct the \emph{feedback factor}, a general quantity, which lumps all the necessary information for efficiency and power analysis of a thermodynamic system. Much of the terminology at the heart of the present work borrows from control theory but, to avoid confusion, it should be noticed that nothing in our approach is entirely specific to control theory. The article is organized as follows: we first present the basic and general set of equations that underlies our closed loop approach. We then illustrate the analysis with the case study of a thermoelectric system. Discussion and concluding remarks end the article.

\paragraph*{General framework}

\begin{figure}[h]
\includegraphics[scale=0.5]{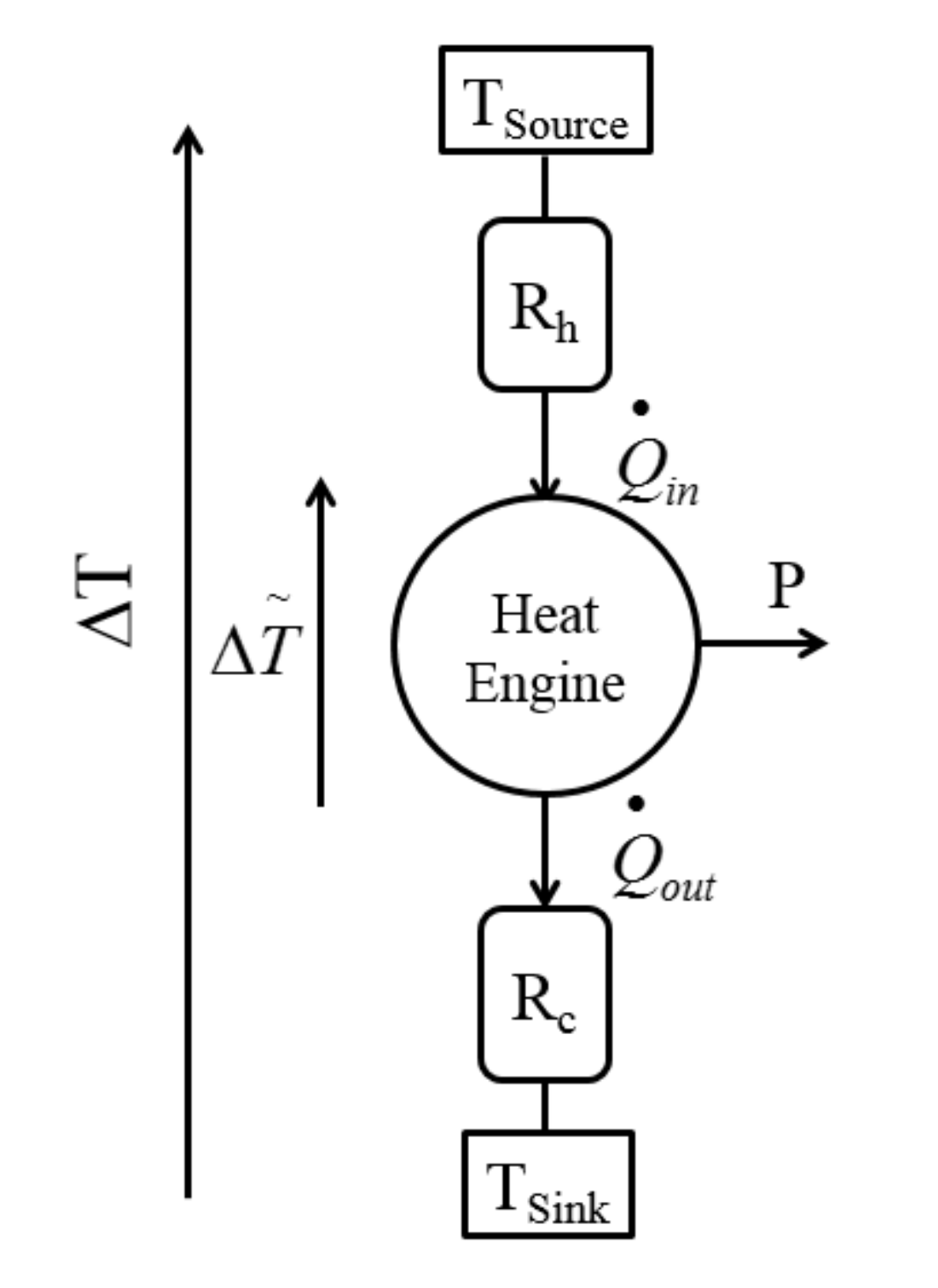}\caption{Nodal description of a thermodynamic system composed of a conversion engine coupled to two constant temperature baths through heat exchangers.}
\label{figure1}
\end{figure}

Consider a system composed of a conversion engine and the two heat exchangers that connect it to two thermal reservoirs at different constant temperatures, as depicted in Fig.~\ref{figure1}. While the temperature difference between the two thermal baths is constant and equal to $\Delta T=T_{\mathrm{source}}-T_{\mathrm{sink}}$, the temperature difference $\Delta\widetilde{T}$ between the two ends of the conversion engine is not fixed as it depends on the working conditions of the engine, and hence on the output power $P$ produced or received by the engine. It should be mentioned that $P$ also accounts for the engine internal dissipation due to friction. Each heat exchanger has a finite thermal resistance, $R_{\mathrm{h}}$ and $R_{\mathrm{c}}$ respectively, and is thus a location of entropy production. The total thermal resistance connected to the device is, 

\begin{equation}
R_{\mathrm{c}}+R_{\mathrm{h}}=R_{\mathrm{\theta}}
\label{Rtheta}
\end{equation}

Concerning the engine itself, we make no assumption on the ideal (endoreversible) or imperfect nature of the heat-to-work conversion. The output power is simply derived from the power budget: $P=\dot {Q}_{\mathrm{{in}}}-\dot{Q}_{\mathrm{{out}}}$, with $\dot{Q}_{\mathrm{in}}$ the heat flux that enters the engine, and $\dot{Q}_{\mathrm{out}}$ the rejected heat flux. 

The thermal budget may be expressed as
 
\begin{equation}
\Delta T=\dot
{Q}_{\mathrm{in}}R_{\mathrm{h}}+\dot{Q}_{\mathrm{out}}R_{\mathrm{c}}+\Delta\widetilde{T} \label{Tbudget}%
\end{equation} 

\noindent The mean heat flux flowing inside the engine is $\dot{Q}_{\mathrm{m}}=\dot{Q}_{\mathrm{out}}+\delta\dot{Q}=\dot{Q}_{\mathrm{in}}-\delta\dot{Q}$, and the power reads $P=\dot{Q}_{\mathrm{in}} - \dot{Q}_{\mathrm{out}} =2\delta\dot{Q}$. Then we get $\Delta T=\left(  \dot{Q}_{\mathrm{m}}+\delta\dot{Q}\right) R_{\mathrm{h}}+\left(  \dot{Q}_{\mathrm{m}}-\delta\dot{Q}\right) R_{\mathrm{c}}+\Delta\widetilde{T}$. Now, introducing $\lambda$ as an adjustable parameter controlling the heat exchanger resistance balance, such that $R_{\mathrm{h}}=\lambda R_{\mathrm{\theta}}$ and $R_{\mathrm{c}}=\left(1-\lambda\right)  R_{\mathrm{\theta}}$, we now rewrite the thermal budget as follows: 
 
\begin{equation}
\Delta T=\left(  2\lambda-1\right)  R_{\mathrm{\theta}}\delta\dot{Q}+R_{\mathrm{\theta}}\dot{Q}_{\mathrm{m}}+\Delta\widetilde{T} \label{Qsigma} 
\end{equation} 

\noindent With no loss of validity and for the sake of simplicity, we may focus on the symmetric case: $\lambda=1/2$, so that $\Delta T=R_{\mathrm{\theta}}\dot{Q}_{\mathrm{m}}+\Delta\widetilde{T}$. We also  define the equivalent thermal resistance of the engine as $R_{\rm th}=\Delta\widetilde{T}/\dot{Q}_{\mathrm{m}}$. It is clear that $R_{\rm th}$ is strongly depending on the specific working conditions of the system, as emphasized below. The thermal analogue of the voltage divider formula \cite{ApertetEPL,ApertetEPLr} gives $\Delta\widetilde{T}=\Delta T~R_{\rm th}/(R_{\rm th}+R_{\theta})$. 

According to irreversible thermodynamics, the general description of a conversion engine requires a set of intensive thermodynamic potentials $(T,X)$ and their conjugate (extensive) currents $(\dot{Q},I_{X})$: $T$ is the temperature variable, and $X$ may be the electrochemical potential, the pressure, or any other intensive variable, depending on the specific system of interest; the conjugate current of $X$, $I_{X}$, may thus be, e.g., the output electrical current, a dynamic torque or a mass velocity. We also define the potential conversion factor and the flux conversion factor, respectively as: 

\begin{eqnarray}
\label{A0X}
A_{0X}&=&\Delta X/\Delta\widetilde{T}\\
\label{A0F}
A_{0F}&=&I_{X}/\Delta\widetilde{T} 
\end{eqnarray}

The system is the locus of the competition between a net produced power and its internal dissipation. Assuming a linear out-of-equilibrium description, this internal dissipation adopts a quadratic form with respect ot the current, i.e. proportional to $I_{X}^{2}$. A fraction of this dissipated power is rejected both on the hot and cold sides of the engine \cite{Apertet2013,Apertet2014}. 

\paragraph*{Linear closed-loop approach}

\begin{figure}[ptb]
\includegraphics[scale=0.4]{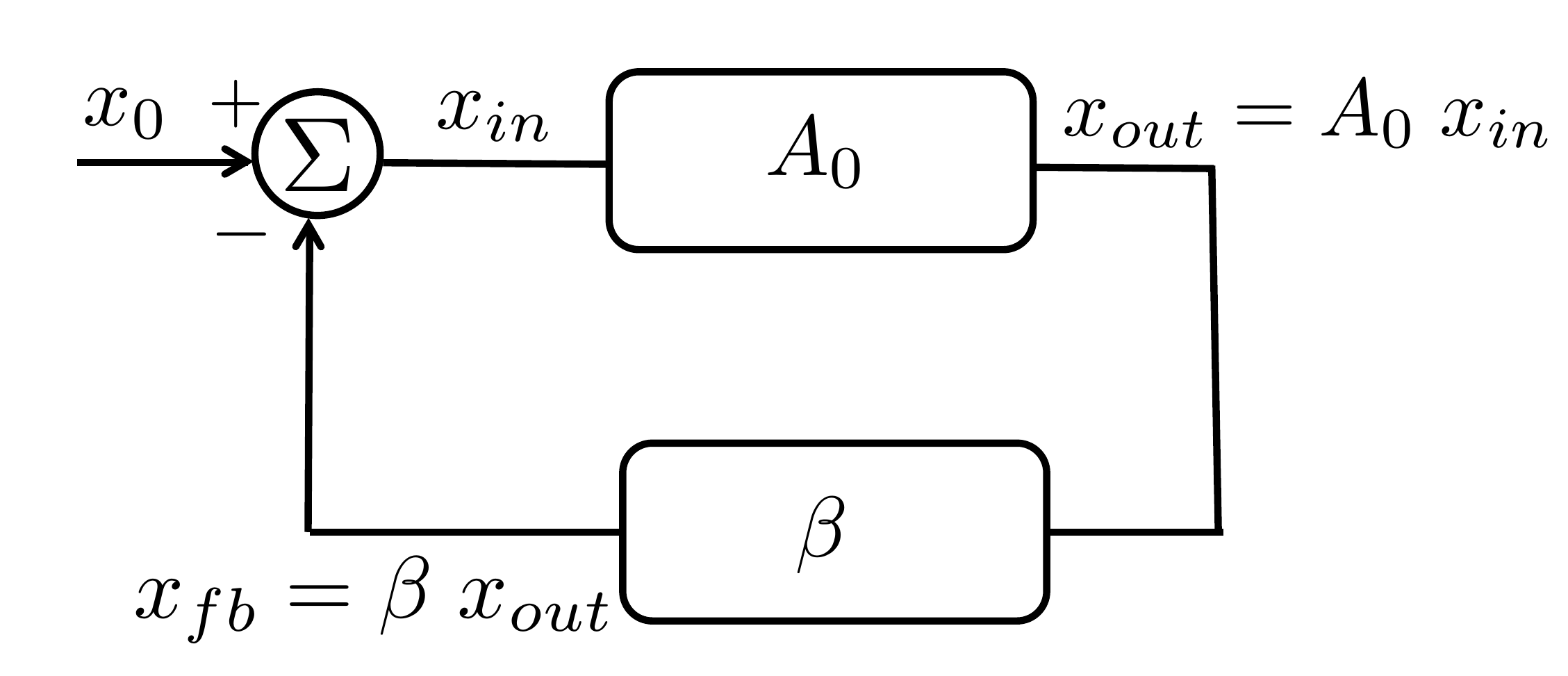}\caption{General description of a linear closed-loop system. The outgoing variable $x_{\rm out}$ results from a multiplication by $A_{0}$ of the entering $x_{\rm in}$, and a feedback contribution of the $\beta$ factor.}\label{figure2}
\end{figure} 

The linear closed-loop description is a common concept in circuit theory \cite{EmDeg}. Under a closed-loop configuration, a system may present feedback effects that involve some modification of the input quantity $x_{\rm in}$, say a voltage, current or power, by the injection of a fraction of the output signal $x_{\rm out}$, as depicted in Fig.~\ref{figure2}. Here, we show that feedback effects also occur in thermodynamic systems because of the presence of the thermal coupling via the heat exchangers to the temperature baths. This means that there exists a feedback effect in thermodynamic systems that emerges from the presence of thermal contacts. \emph{This appears to be a complete change of paradigm because the physical behavior of the system is now mainly driven by its thermal boundary conditions.} 

\begin{figure}[h]
\includegraphics[scale=0.32]{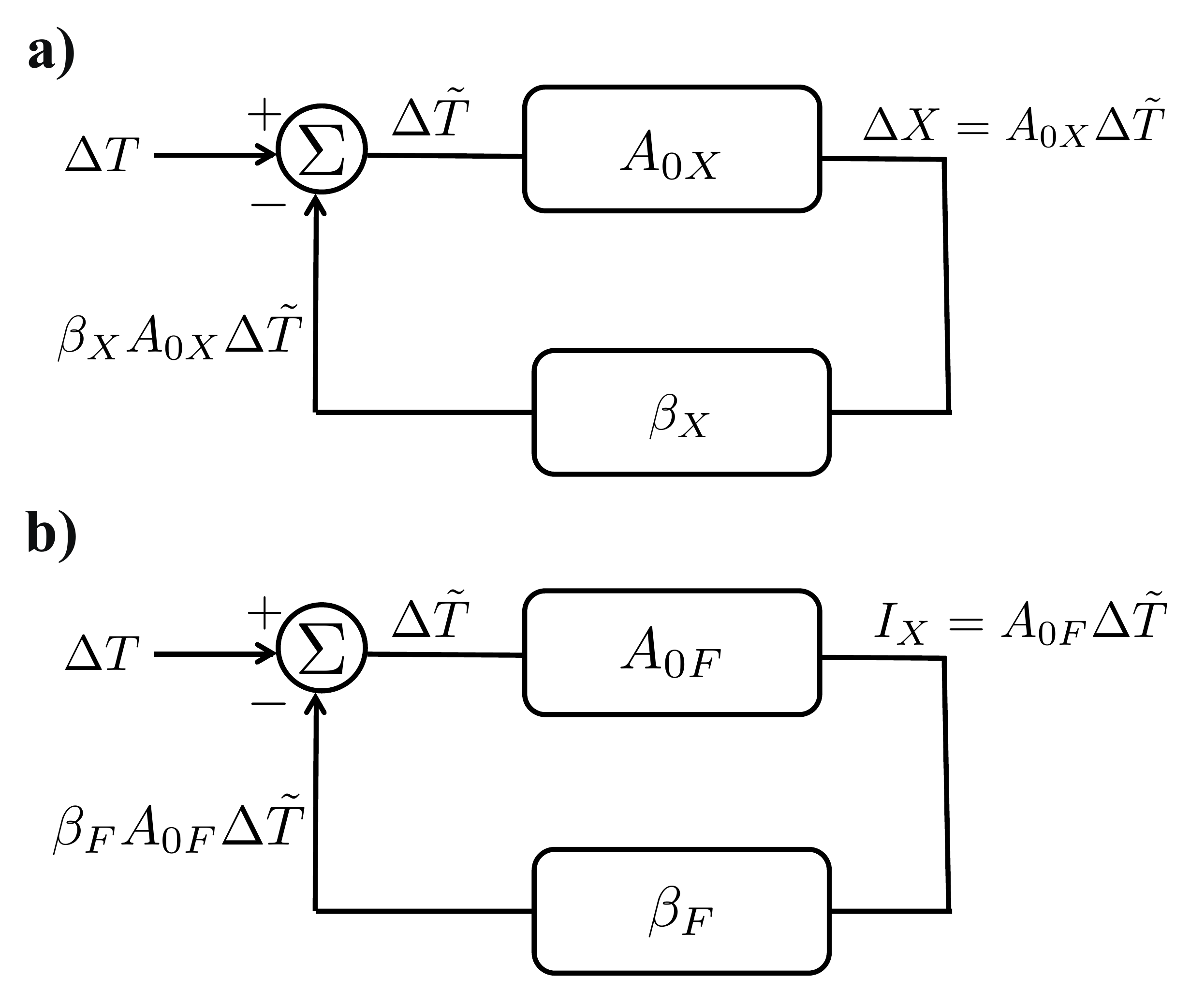}\caption{Conversion engine as a closed-loop system in a) the temperature-potential feedback configuration, and b) the temperature-flux feedback configuration.}
\label{figure3}
\end{figure}

To describe this feedback effect, we first consider a generic closed-loop system. The system presents a direct gain, called open-loop gain $A_{0}$, and a feedback term $\beta$. It follows that the closed-loop gain is simply defined as $A_{\rm cl}=x_{\rm out}/x_{0}=A_{0}/(1+A_{0}\beta)$. In control theory the configuration $A_{0}\beta=-1$ is known to be the oscillating condition for the linear system, which is considered below. It should be noticed that $x_{\rm in}$ and $x_{\rm out}$ may be of different nature, hence have different dimensions. We now proceed with the analysis of the thermodynamic system depicted in Fig.~\ref{figure3}, considering the feedback between the potentials, i.e. the \emph{temperature-potential feedback}, where both $x_{\rm in}$ and $x_{\rm out}$ are potentials. From Eq. \eqref{Qsigma} and the expression for $R_{\rm th}$, the closed-loop gain may be obtained in the form: 

\begin{align}
\frac{1}{A_{\rm cl}}=\frac{x_{0}}{x_{\rm out}}=\frac{\Delta T}{A_{0X}\Delta\widetilde{T}}=\frac{1+R_{\theta}/R_{\rm th}}{A_{0X}}=\frac{1+A_{0X}\beta_{X}}{A_{0X}}. 
\end{align}

\noindent from which the gain-feedback product is identified as:

\begin{align}
A_{0X}\beta_{X}=\frac{R_{\theta}}{R_{\rm th}}
\end{align}

\noindent Then, the closed-loop gain simply becomes $A_{\rm cl}=A_{0X}R_{\rm th}/(R_{\theta}+R_{\rm th})$. The \textit{temperature-flux feedback} is obtained in a similar fashion, where $x_{\rm in}$ is a temperature difference and $x_{\rm out}$ is a flux. The corresponding open-loop gain $A_{0F}$ together with the feedback factor $\beta_{F}=R_{\mathrm{\theta}}/(A_{0F}R_{\rm th})=A_{0X}\beta_{X}/A_{0F}$ then yield 

\begin{align}
A_{0F}\beta_{F}=A_{0X}\beta_{X}=A_0\beta=\frac{R_{\theta}}{R_{\rm th}} 
\end{align}

\noindent which can also be expressed as the ratio of the temperature differences: 

\begin{align}
\frac{\Delta\widetilde{T}}{\Delta T}=\frac{1}{1+A_0\beta} 
\end{align}

For a given temperature difference between the reservoirs, and for a given open-loop gain, the different working points of the system are determined by the different values that the gain-feedback product $A_0\beta$ may take. The ratio $\Delta\widetilde{T}/\Delta T$ allows for all the possible working configurations of a thermal engine under hybrid (potential and flux) thermal boundary conditions. One may try to have $\Delta\widetilde{T}$ as large as possible to maximize efficiency or, conversely, have $\Delta\widetilde{T}$ as small as possible in order to maximize the heat flux intensity. Note that the product $A_0\beta$ includes both the engine response through $R_{\rm th}$ and the boundary conditions fixed by the coupling to the reservoirs through $R_{\theta}$. In other words, \emph{the product $A_0\beta$ defines the working mode of the overall system that includes both the conversion engine and the associated thermal boundary conditions.} With $A_{\rm cl}=A_{0}/(1+A_{0}\beta)$, the closed-loop system acts as an open loop when $A_0\beta=0$ i.e. $R_{\theta}=0$. This condition corresponds to a perfect thermal coupling to the fixed temperature reservoirs. In agreement with control theory, in this configuration, there is no feedback effect wich is expected for systems under Dirichlet (i.e. potential) boundary conditions. 

To briefly illustrate the general results derived so far, it is instructive to consider the simple case of a heat engine operating a transmission shaft under two configurations: the \emph{zero-load} configuration, which corresponds to the situation when the engine operates with no load; and the \emph{blocking-load}, which corresponds to the case where the shaft rotation is completely blocked by the load. The zero-load configuration is simply characterized by the potential conversion factor under the zero-flux condition: $A_{0X}^{\star}=\Delta X/\Delta\widetilde{T}$, while the blocking-load configuration amounts to having $A_{0X}=0$, and a finite feedback parameter $\beta_F^{\star}$, which accounts for the engine's internal friction forces. Table~\ref{Table1} summarizes the results for the potential conversion and flux conversion factors, the feedback parameters and the closed loop-gain for this example. 

\begin{table}
\caption{\label{Table1}Comparison of the potential and flux conversion and feedback factors, and closed-loop gain under the zero-load and blocking-load conditions.}
\begin{tabular}{l| c c c c c}
\hline\hline
~&~&~&~&~&~\\
conversion, feedback, and gain& $A_{0X}$ & $A_{0F}$ & $\beta_{X}$ & $\beta_{F}$ & $A_{\rm cl}$\\
~&~&~&~&~&~\\
zero load & $A_{0X}^{\star}$ & $0$ & $\frac{R_{\theta}}{\alpha R_{\rm th}}$ & $\infty$ & $\frac{A_{0X}^{\star}}{1+\frac{R_{\theta}}{R_{\rm th}}}$\\
~&~&~&~&~&~\\
blocking load & $0$ & $\frac{\alpha}{\beta_{F}^{\star}}\frac{R_{\theta}}{R_{\rm th}}$ & $\infty$ & $\beta_{F}^{\star}$ & $0$\\
~&~&~&~&~&~\\
\hline\hline
\end{tabular}
\end{table}

We may now proceed and extend our analysis to the \emph{temperature-power feedback}. Using Eqs.~\eqref{A0X} and \eqref{A0F}, the output power is thus defined as: 

\begin{align}
P=\Delta X I_{X}=A_{0X}A_{0F}\left(\Delta\widetilde{T}\right)^2 =A_{0X}A_{0F}\left(\frac{\Delta T}{1+A_0\beta}\right)^2
\end{align}

\noindent Following the same procedure, one may obtain the general expression for the efficiency $\eta$ as:

\begin{align}
\eta=\frac{P}{\dot{Q}_{\mathrm{in}}}=
\frac{A_{0X}A_{0F}\Delta T}{R_{\rm th}}\left(\frac{1}{1+A_0\beta}\right)^3
\end{align}

\noindent assuming that $\dot{Q}_{\mathrm{in}}\approx(R_{\theta}+R_{\rm th})\Delta T$. Notice that, contrary to $P$, the efficiency depends on both $A_0\beta$ and $R_{\rm th}$. The oscillating configuration is obtained when $R_{\mathrm{\theta}}/R_{\rm th}=-1=e^{i\pi}$. However, in most practical systems, the oscillating condition seldom occurs, since the ratio $R_{\mathrm{\theta}}/R_{\rm th}$ is usually real and positive. Only thermo-acoustic system may fulfilled this conditions since impedance of the heat re-generator of such system is complex and thus contains a reactive part \cite{acoustic}.

\begin{figure}[h]
\includegraphics[scale=0.55]{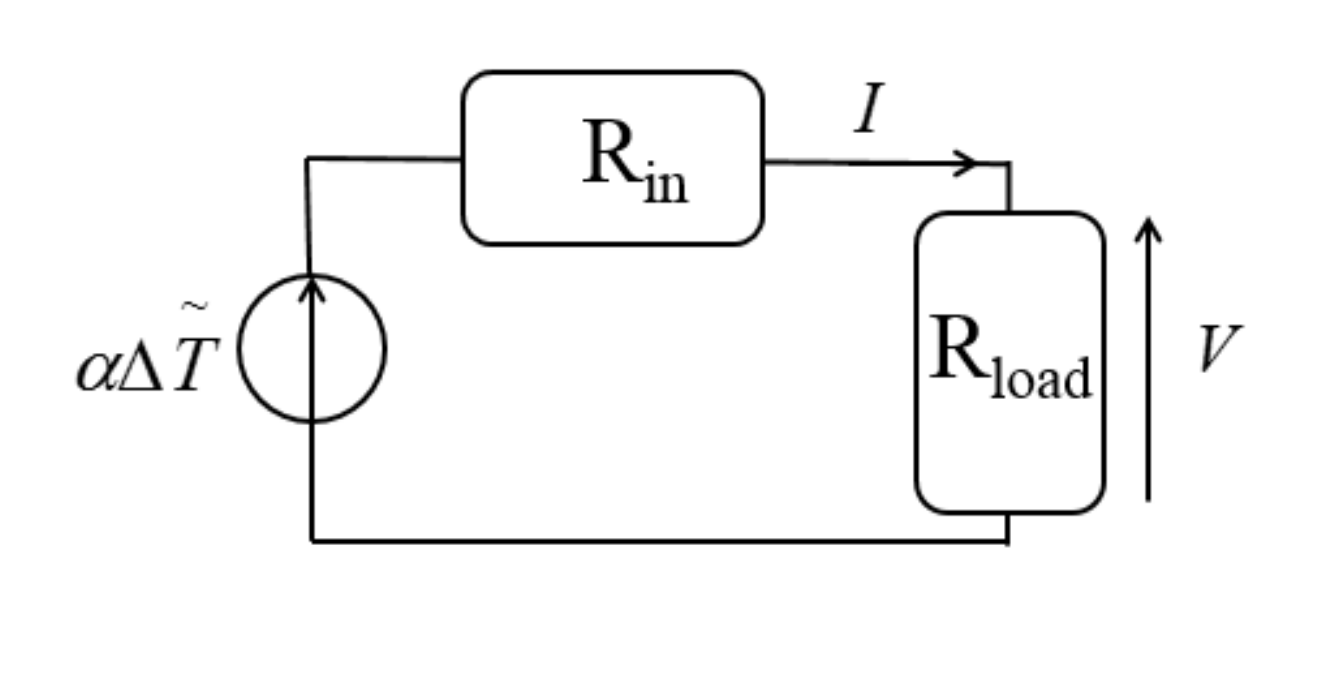}\caption{Basic electrical description of a thermoelectric system.}\label{figure4}
\end{figure}

\paragraph*{Application to a thermoelectric generator}
Thermoelectric systems may operate as generators or heat pumps, by conversion of a heat current into electric power and conversely. Consider a thermoelectric generator with an internal electric resistance $R_{\mathrm{{in}}}$, and a simple resistive load $R_{\rm {load}}$ to which it is connected to (Fig.~\ref{figure4}).

The coupling between the temperature $T$ and electrochemical potential $\mu$ is determined by the Seebeck coefficient $\alpha$ under zero electrical current \cite{Apertet2016}: 

\begin{align}
\alpha=\frac{\Delta(\mu/e)}{\Delta T} 
\end{align}

\noindent with $e$ being the electron charge. These conversion engines use the conduction electrons as the thermodynamic working fluid. This working fluid is characterized by the so-called figure of merit \cite{Ioffe}: $Z\overline{T} =\alpha^{2}(R_{\rm th0}/R_{\rm in})\overline{T}$, where $\overline{T}$ is the average temperature across the engine and $R_{\rm th0}$ denotes its thermal resistance at zero electrical current.

The output electrical power is $P=VI$ with $I$ being the electrical current and $V$ being the voltage given by $V=\alpha\Delta\widetilde{T}R_{\rm load}/(R_{\rm in}+R_{\rm load})=R_{\mathrm{{load}}}I$, with $\Delta\widetilde{T}$ given below Eq.~\eqref{Qsigma}. So, adapting Eqs.~\eqref{A0X} and \eqref{A0F} to the present case, we simply get 
$A_{0X}=A_{0V}=V/\Delta\widetilde{T}$ and $A_{0F}=A_{0I}=I/\Delta\widetilde{T}$, and the closed-loop parameters for the considered thermoelectric system are: 

\begin{eqnarray}
\label{A0XTE}
A_{0V} &=&A_{0I}R_{\rm load}=\alpha\frac{\rm R_{\rm load}}{R_{\rm in}+R_{\rm load}}\\ 
\label{betaXTE}
\beta_{V} &=&\frac{\beta_{I}}{R_{\rm load}}=\frac{R_{\theta}}{A_{0V}R_{\rm th}}=\frac{R_{\theta}}{R_{\rm th}}\frac{R_{\rm in}+R_{\rm load}}{\alpha R_{\rm load}}
\end{eqnarray} 

\begin{figure}[ptb]
\includegraphics[scale=0.3]{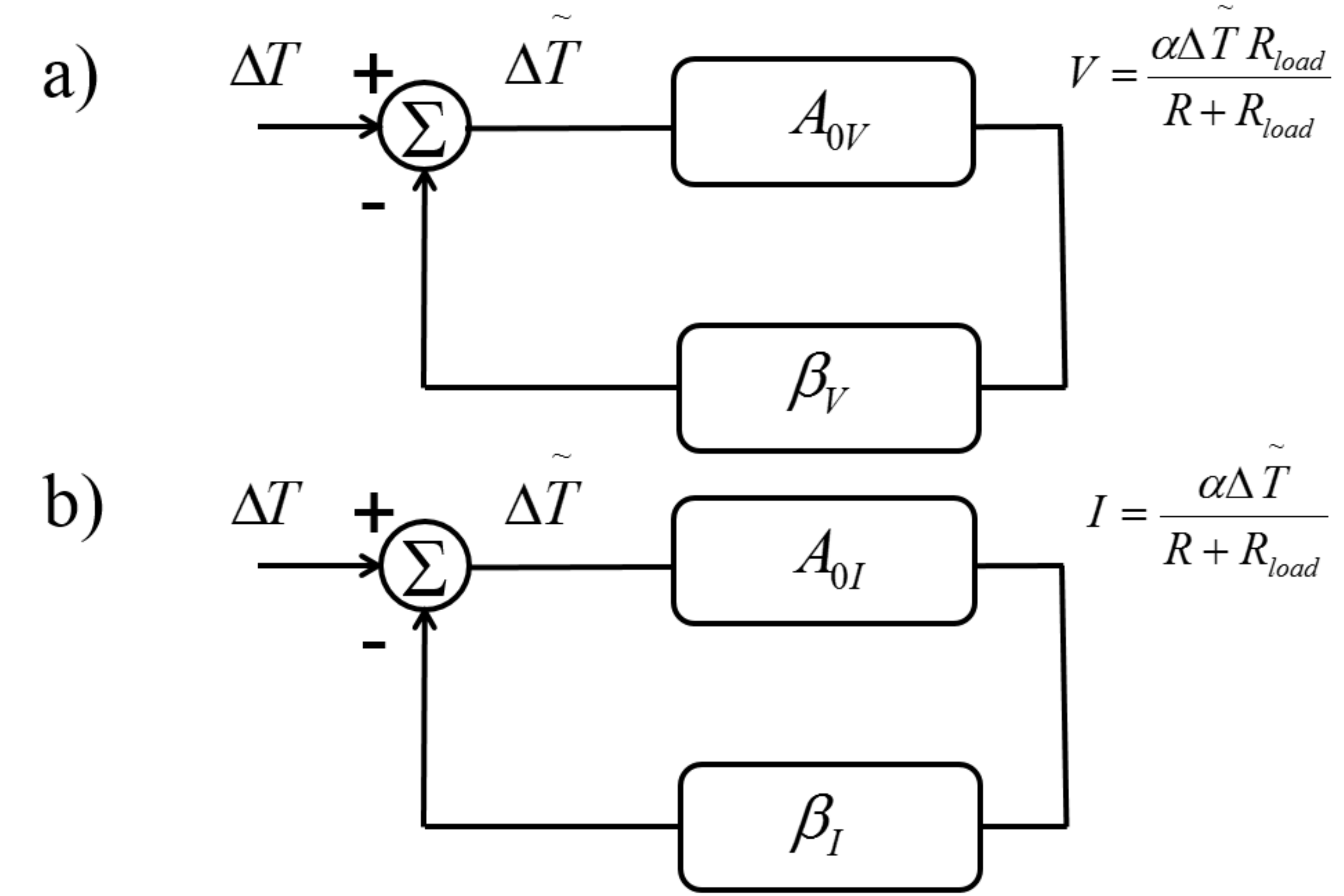}\caption{Thermoelectric system as a closed-loop system, with a) temperature-voltage loop, and b) temperature-current loop.}
\label{figure5} 
\end{figure}

\noindent The corresponding closed-loop description is depicted on Fig.~\ref{figure5}. It is worth noticing that the open loop gain is only affected by the electrical working conditions, hence $R_{\rm load}$ in this case. In particular, in the open voltage configuration, $R_{\rm load}\to\infty$, $A_{0V}$ actually coincides with the Seebeck coefficient, which is defined under zero current conditions. Then, $A_{0V}$ appears to be the extension of the Seebeck coefficient definition for non zero current configurations. 

By contrast to $A_{0V}$, the feedback term is determined by both the thermal and electrical working conditions. This confirms the essential role of the question relative to accurately define the thermal boundary conditions for a working thermal engine \cite{CurzonAhlborn,Ouerdane2015a}. Under Dirichlet thermal boundary conditions, $R_{\theta}=0$, the temperature difference across the engine is fixed and there is no feedback as $\beta_{V}=\beta_{I}=0$. In this case, $\Delta\widetilde{T}=\Delta T$, as expected for Dirichlet conditions. Under Neumann conditions, $R_{\theta}\to\infty$, the feedback terms go to infinity. This divergence has no physical consequence because the input temperature difference and, consequently, the open voltage $\alpha\Delta\widetilde{T}$ vanish in this case. If we now consider the electrical working conditions, one can see that under open-circuit conditions $(R_{\rm load}\rightarrow\infty)$ the open-loop gain is simply equal to $\alpha$ and the current open-loop gain vanishes, while the voltage feedback saturates to $\beta_{V}=R_{\theta}/R_{\rm th}$. The current feedback diverges. In this configuration, the system does not produce any output power but only maintains an output voltage. In contrast, under short-circuit condition, which is also a zero output power configuration, $(R_{\rm load}=0)$, the voltage feedback diverges and the current feedback saturates to $R_{\theta}R_{\rm in}/(R_{th}\alpha)$. The voltage open-loop gain vanishes and the current open-loop gain saturates to $\alpha/R_{\rm in}$. In this case, the output voltage is zero and the output current is maximal. But, due to the absence of any output voltage, the produced electrical power is \emph{totally} reinjected into the system and fully dissipated. One may argue that there is no power produced neither outside nor inside the system in this configuration; however, this view is not correct, because i) $\alpha$ is built from the equation of state of the electronic fluid, and is never zero; and ii) because in this case $\Delta\widetilde{T}$ does not go to zero, so an open voltage exists. This configuration is similar to the case where a short-circuited battery dissipates all its power by \emph{internal} Joule effect. Note that this situation may also occur in the so-called Rayleigh-Bernard systems \cite{Rayleigh-Benard}. The different configurations are summarized in Table \ref{Table2}.

\begin{table}
\caption{\label{Table2}Comparison of the potential and flux conversion and feedback factors, and closed-loop gain of a thermoelectric generator under the open-circuit and short-circuit conditions.}
\begin{tabular}
{l|c c c c c}\hline\hline 
~&~&~&~&~&~\\
conversion, feedback, and gain& $A_{0V}$ & $A_{0I}$ & $\beta_{V}$ & $\beta_{I}$ & $A_{\rm cl}$\\
~&~&~&~&~&~\\
open-circuit & $\alpha$ & $0$ & $\frac{R_{\theta}}{\alpha R_{\rm th}}$ & $\infty$ & $\frac{\alpha}{1+\frac{R_{\theta}}{R_{\rm th}}}$\\
~&~&~&~&~&~\\
short-circuit & $0$ & $\frac{\alpha}{R_{\rm in}}$ & $\infty$ & $\frac{R_{\theta}R_{\rm in}}{\alpha R_{\rm th}}$ & $0$\\
~&~&~&~&~&~\\
maximal power & $\frac{\alpha}{2}$ & $\frac{\alpha}{2R_{\rm in}}$ &$\frac{2R_{\theta}}{\alpha R_{\rm th}}$ & $\frac{2R_{\theta}R_{\rm in}}{\alpha R_{\rm th}}$ & $\frac{\frac{\alpha}{2}}{1+\frac{R_{\theta}}{R_{\rm th}}}$\\
~&~&~&~&~&~\\
\hline\hline
\end{tabular}
\end{table}

This analysis can be extended considering the derivation of the equivalent thermal resistance $R_{\rm th}$ of a thermoelectric engine \cite{ApertetEPL,ApertetEPLr,ApertetJPC}, in which case we have:

\begin{equation}
R_{\rm th}=R_{\rm th0}/(1+\sigma_{\rm P}) 
\label{Rth}
\end{equation}

\noindent with $\sigma_{\rm P}=R_{\rm in}/(R_{\rm in}+R_{\rm load})Z\overline{T}$ being the so-called thermoelectric Prandtl number \cite{ApertetJPC}. The gain $A_{0V}$ is not modified by this expression but the feedback term now reads:

\begin{equation}
\beta_{V}=\frac{\beta_{I}}{R_{\rm load}}=R_{\theta}\frac{R_{\rm load}+R_{\rm in}(1+Z\overline{T})}{\alpha R_{\rm th0}R_{\rm load}} 
\end{equation}

\begin{figure}[!rh]
\includegraphics[scale=0.4]{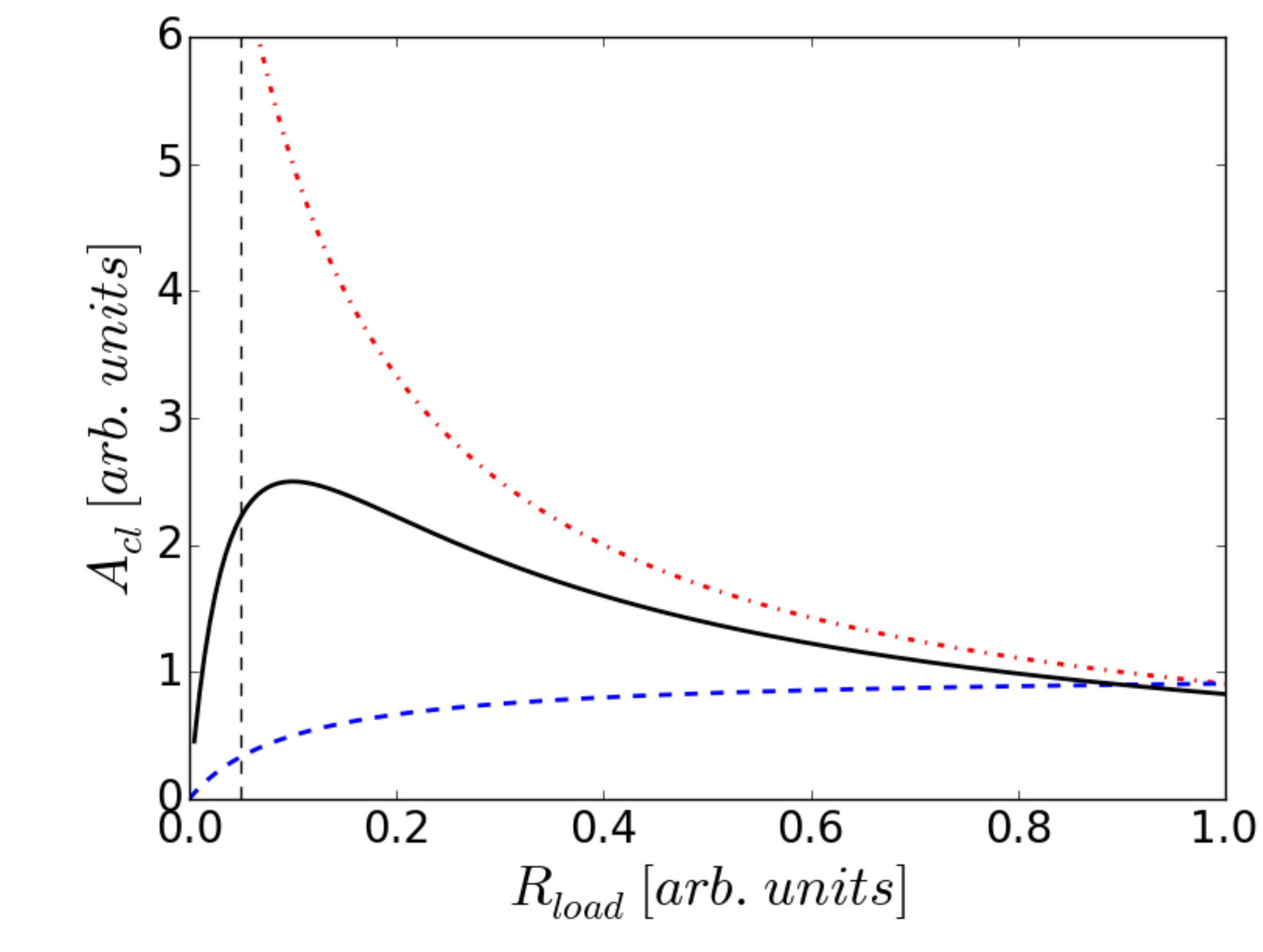}\caption{Voltage (dashed/blue), current (dot dashed/red) and power (solid/black) closed-loop gain. Note that the location of the maximum power does not coincide with the traditional assumption $R_{load}=R_{\rm in}$ (vertical dashed/black line on the left side of the panel).} 
\label{figure6} 
\end{figure} 

The respective responses of voltage, current and power for the thermoelectric system and for different working points are reported in Fig.~\ref{figure6}. Both voltage and current closed-loop are monotonous functions of the output load. For the power, the closed-loop gain results from the competition between the voltage and current gains. In order to derive the expression of the maximum output power one has to satisfy the electric impedance matching condition: $R_{\rm load}^{\rm (max)}=R_{\rm in}+R_{\rm TE}$, which is the load value for a maximal output power, with $R_{\rm TE}=R_{\rm in}\alpha^2\overline{T}/(1/R_{\rm th0}+1/R_{\theta})$ being the so-called \textit{thermoelectric resistance} which has also been theoretically and experimentally reported from impedance spectroscopy analysis \cite{ApertetEPL,ApertetEPLr,Downey2007}. Surprisingly this result slightly differs from the traditional condition $R_{\rm load}=R_{\rm in}$, and only reduces to this expression under the Dirichlet configuration where no feedback is present. This confirms that the traditional description and optimization of thermoelectric systems, which only consider the latter condition for maximal output power, systematically neglects the feedback contributions. 

\paragraph*{Oscillating configuration}
As previously mentioned, the closed-loop gain may diverge if the condition $A_0\beta=R_{\theta}/R_{\rm th}=e^{i\pi}$ is fulfilled. Assuming complex thermal impedances for the system, the previous ratio may then present a phase shift. This phase rotation effect is at the origin of electronic oscillators like Wien bridges, which act as narrow-band amplifiers. Indeed, these structures are very efficient electron current filters. One relevant question that arises then concerns the possibility or not to \textit{access oscillating conditions for a thermoelectric system}. Although not reported yet in thermoelectric systems, such spontaneous oscillations have already been observed in other thermal systems like thermo-acoustic engines, and also other thermodynamic systems like Teorell oscillators \cite{Teorell1959}. 

To proceed with this question, let us now extend the thermal resistances Eqs. \eqref{Rtheta} and \eqref{Rth} to complex thermal impedances, ${\mathcal Z}_{\rm th}$ and ${\mathcal Z}_{\theta}$. Considering both the in-phase and out-of-phase terms we get:

\begin{eqnarray}
\label{zth}
{\mathcal Z}_{\rm th} &=& \frac{{R_{\rm th0}}}{\left(1+\frac{R_{\rm in}}{R_{\rm in}+R_{\rm load}}Z(i\omega)\overline{T}\right)}\\
\label{zteta}
{\mathcal Z}_{\theta} &=& \frac{R_{\theta}}{1+i\frac{\omega}{\omega_{\theta}}} 
\end{eqnarray}

\noindent where $1/\omega_{\theta}$ is the thermal time constant of the heat exchangers, and $Z(i\omega)\overline{T}$ the frequency-dependent figure of merit. The derivation of Eqs. \eqref{zth} and \eqref{zteta} follows from classical transport equations in the frequency domain \cite{transient,Ezzahri2014}. The transport coefficients: electric conductivity $\sigma$, thermal conductivity, $\kappa$, and Seebeck coefficient, $\alpha$, thus read \cite{Ezzahri2014}: 

\begin{eqnarray}
\sigma(i\omega)  &  = & \frac{\sigma_{0}}{1-i\frac{\omega}{\omega_{\sigma}}}\\
\kappa(i\omega)  &  = & \frac{\kappa_{0}}{1-i\frac{\omega}{\omega_{\kappa}}}\\
\alpha(i\omega)  &  = & \frac{\alpha_{0}}{1-i\frac{\omega}{\omega_{\alpha}}} 
\end{eqnarray} 

\noindent $\omega_{\sigma}$, $\omega_{\kappa}$, and $\omega_{\alpha}$ are the low-pass cut-off frequencies of each transport coefficient. 

\begin{figure}[ptb]
\includegraphics[scale=0.4]{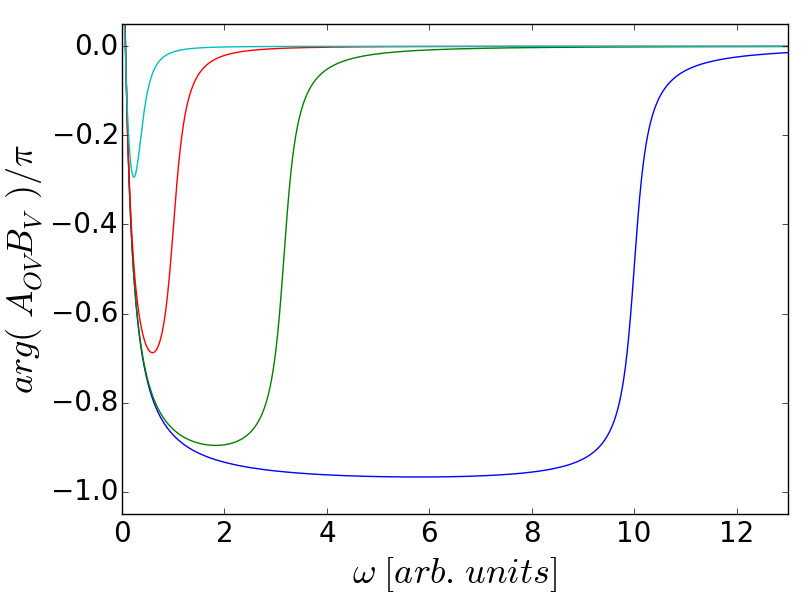}
\caption{$\omega$-dependence of $\arg(A_{0X} \beta_{X})/\pi$ with varying $\omega_{\kappa}$ in the range 10$^{-2}$ (cyan) to 10$^{-5}$ (blue), $\omega_{\alpha}$ and $\omega_{\sigma}$ being fixed to 0.1.}
\label{figure7}
\end{figure}

\begin{figure}[ptb]
\includegraphics[scale=0.4]{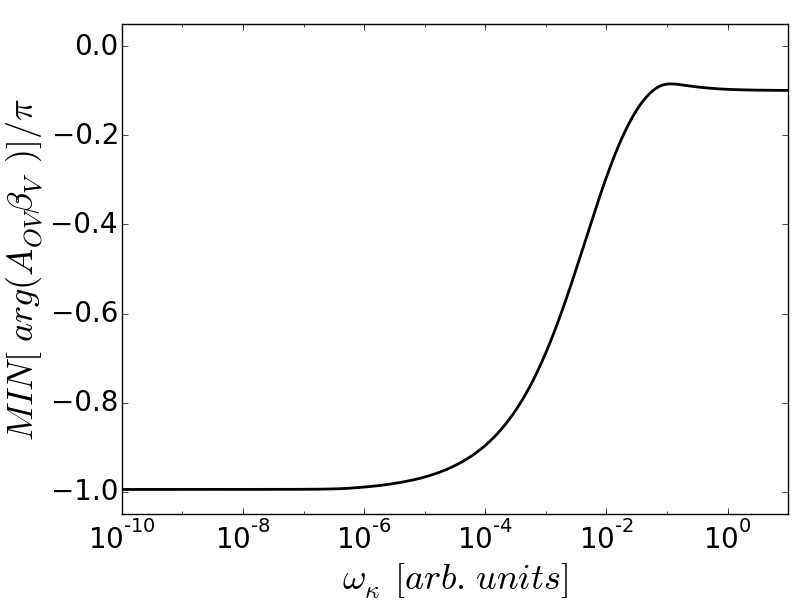}
\caption{Minimum value of $\arg(A_{0X} \beta_{X})/\pi$ as a function of $\omega_{\kappa}$.}
\label{figure8}
\end{figure}

Substituting these expressions into Eqs.~\eqref{A0XTE} and \eqref{betaXTE} with $Z(i\omega)\overline{T}=\sigma(i\omega)\alpha^{2}(i\omega)\overline{T}/\kappa(i\omega)$, we obtain $A_{0V}\beta_{V}(j\omega)$, the obtained final expression is: 

\begin{equation}
A_{0V}\beta_V(i\omega)=\frac{R_{\theta}}{R_{\rm th0}}\left(  1+\frac{R_{\rm in} }{R_{\rm in}+R_{\rm load}}Z\overline{T}(i\omega)\right)  \label{ABeta} 
\end{equation}

Numerical results are shown in Fig.~\ref{figure7}. The total phase shift is found to asymptotically approach $-\pi$ if and only if the modulus $|Z\overline{T}|$ diverges. In addition, this only occurs for vanishing frequencies as shown on Fig.~\ref{figure8} This condition is fulfilled only if the thermodynamic working fluid is ideal (no internal dissipation) and if the engine works in the fully reversible mode. The simulation has been performed assuming a constant resistive load. Let us now consider the influence of a complex load ${\mathcal Z}_{\rm load}(i\omega)$ instead of a purely real load $R_{\rm load}$. The non-zero imaginary part of this complex load may influence the response of the system if and only if $|{\mathcal Z}_{\rm load}(i\omega)| \gg R_{\rm in}$. This is consistent with the fact that the occurrence of oscillations can only exist in low dissipative engines, where the internal resistance is sufficiently small.

Assuming both a small internal resistance $R_{\rm in}$ and a large value for $|Z\overline{T}|$, the expression becomes $A_{0V}\beta_V(i\omega)\propto\frac{R_{\theta}}{R_{\rm th0}R_{\rm load}}\left(\frac{\alpha^{2}(i\omega)\overline{T}}{\kappa(i\omega)}\right)$. Since the $\omega$ and $2\omega$ signals are orthogonal components in the frequency domain, the total phase shift never benefits from the addition of the $\alpha^{2}(i\omega)$ contribution. Then the desired $\pi$ phase shift can only be provided by the product ${\mathcal Z}_{\rm load}(i\omega)\kappa(i\omega)$. If the ${\mathcal Z}_{\rm load}(i\omega)$ frequency-dependence is of the first order, then the maximal phase shift giving the oscillating condition becomes $\mbox{Arg}\left[{\mathcal Z}_{\rm load}(i\omega)\kappa(i\omega)\right]=\pi$, which may only occur at very low frequencies, if any. In other words, the oscillating condition for a macroscopic thermoelectric system requires a complex load with a ${\mathcal Z}_{\rm load}(i\omega)$ frequency-dependence exponent strictly larger than one, like e.g. a second-order resonating load.  

\paragraph*{Application to mesoscopic thermoelectric systems}
Considering a mesoscopic system, the main difference occurs from the coupling parameters that cannot be described as simple resistive terms, but include a capacitance component in series. With a kinetic integral approach, it can be shown that the heat flow is driven by the carrier transport at energy above the Fermi level. 

In this condition, the thermal coupling can be expressed as ${\mathcal Z}_{\theta}(i\omega)$ instead of $R_{\theta}$, modeling the series resistance/capacitance term, the admittance of which is given by the so-called quantum capacitance as proposed by B\"uttiker \cite{Buttiker1993}. At low frequencies, the response is mostly resistive and yields no significant phase shift. At high frequencies, the phase shift induced by ${\mathcal Z}_{\theta}(i\omega)$ is close to $\pi/2$, but the modulus also vanishes, and so does the product $A_{0V}\beta_V(i\omega)$. In terms of frequency response behavior, no significant difference between a macroscopic and a mesoscopic system can really be observed. For the two of them, the oscillating conditions can be fulfilled under very strict conditions for the complex load ${\mathcal Z}_{\rm load}(i\omega)$. Nevertheless, these oscillations may possibly be obtained using the response of mesoscopic channels under periodic driving \cite{Zhou2015,Ludovico2015}. 

The system may be tuned in order to work at the resonant frequency. This driving mode can be theoretically considered using the Floquet expansion, which permits the analysis of the transport response in periodically unsteady modes. This paves the way to the analysis of possible working modes where the energy conversion would be obtained in a very narrow-banded range, including the electron filtering effect favorable for thermoelectric conversion \cite{Mahan1996,Humphrey2002,Abbout2013}.

\paragraph*{Discussion and concluding remarks} 
Looking for ways to generalize thermodynamics in view of providing a unified framework for the study of a wide range of processes governing the operation of heat engines, be they of mechanical, thermoelectric, or chemical nature, is a challenging yet rewarding endeavour \cite{Bordoni2013}. The original approach proposed in this article is a contribution to this effort. Though borrowing from the terminology of control theory, it is mostly based on diagrammatic reasoning. Through this approach, we obtained compact analytical expressions for the $A_{0}\beta=R_{\theta}/R_{\rm th}$ ratio. In particular, we recover the expression for the so-called thermoelectric resistance, already introduced by other theoretical and experimental works. This feedback analysis was developed assuming that the working conditions of the considered thermodynamic systems are governed by hybrid boundary conditions, where neither the fluxes nor the potentials are imposed. The present work can be extended to unsteady working modes as long as the overall response is based on linear response theory. Also, from a general theoretical point of view, the question of maximal or minimal dissipation production for an out-of-equilibrium system  still remains an open question. Analyzing the boundary conditions of such a system may provide a partial answer. 

\paragraph*{Acknowledgments} We are pleased to thank Dr. Y. Apertet for his careful and critical reading of the manuscript.


\begin{thebibliography}{99}
\bibitem{EmDeg} One classical example is that of the so-called ``emitter degeneration'' observed in the bipolar transistor common emitter configuration. In this case, the presence of a coupling resistance to the source, introduces the feedback effect.
\end{thebibliography}
\end{document}